\input harvmac

\overfullrule=0pt
\parindent=0pt


\def\al {\alpha'}

\def\NSNS{{$NS\otimes NS$}}
\def\RR{{$R\otimes R$}}
\def\ZZ {{\bf Z}}

\def\det{\hbox{\rm det}}
\def\Sl#1{Sl(#1,\ZZ)}
\def\G(#1){\Gamma(#1)}
\def\al{\alpha'}

\def\C|#1{{\cal #1}}
\def\(#1#2){(\zeta_#1\cdot\zeta_#2)}


\def\xxx#1 {{hep-th/#1}}
\def\lr { \lref}
\def\npb#1(#2)#3 { Nucl. Phys. {\bf B#1} (#2) #3 }
\def\rep#1(#2)#3 { Phys. Rept.{\bf #1} (#2) #3 }
\def\plb#1(#2)#3{Phys. Lett. {\bf #1B} (#2) #3}
\def\prl#1(#2)#3{Phys. Rev. Lett.{\bf #1} (#2) #3}
\def\physrev#1(#2)#3{Phys. Rev. {\bf D#1} (#2) #3}
\def\ap#1(#2)#3{Ann. Phys. {\bf #1} (#2) #3}
\def\rmp#1(#2)#3{Rev. Mod. Phys. {\bf #1} (#2) #3}
\def\cmp#1(#2)#3{Comm. Math. Phys. {\bf #1} (#2) #3}
\def\mpl#1(#2)#3{Mod. Phys. Lett. {\bf #1} (#2) #3}
\def\ijmp#1(#2)#3{Int. J. Mod. Phys. {\bf A#1} (#2) #3}

\def\Im{\rm Im}

\def\tr{{\rm tr}}

\def\calt{{\cal T}}
\def\Rfour{t_8t_8R^4}

\parindent 25pt
\overfullrule=0pt
\tolerance=10000

\sequentialequations



\lr\aspinwall{P. Aspinwall, {\it Some Relationships Between
Dualities in String
Theory } Trieste 1995, Proceedings, S-duality and mirror symmetry, Nucl.
Phys. Proc. {\bf 46} (1996) 30, \xxx9508154}
\lr\schwarz{J.H. Schwarz, {\it An $\Sl2$ multiplet of type IIb
superstrings}, \xxx9508143
\plb360(1995)13.}
\lr\gs{M.B. Green and J.H. Schwarz, {\it Supersymmetrical string
theories}, \plb109(1982)444.}
\lr\dkl{L.~Dixon, V.~Kaplunovsky and J.~Louis, {\it Moduli
dependence of String
loop corrections to gauge coupling constants }, \npb355(1991)649}
\lr\gg{M.B.~Green and M.~Gutperle, {\it Effects of D-instantons},
\xxx9701093.}
\lr\ggII{M.B.~Green and M.~Gutperle, {\it Configurations of two
D-instantons},
\xxx9612127.}
\lr\Tdual {A.~Giveon, M.~Porrati and E.~Rabinovici, {\it Target Space
Duality in String Theory }, \xxx9401139, \rep244(1994)77.}
\lr\sch{C.~Schmidhuber {\it D-brane actions }, \xxx9601003,
\npb467(1996)146. }
\lr\hulltownsend{C.M.~Hull and P.K.~Townsend, {\it Unity of Superstring
Dualities }, \xxx9410167, \npb438(1995)109.}
\lr\mth {J.H.~Schwarz, {\it Lectures on Superstring and M-theory
dualities},
  \xxx9607201.}
\lr\greenbrink{M.~Green , J.H.~Schwarz and L.~Brink, {\it N=4 Yang-Mills
and N=8 supergravity as limits of string theories}, \npb198(1982)474. }
\lr\aha {O.~Aharony, {\it String theory dualities from M theory},
\xxx9604103,
\npb476(1996)470.}
\lr\dpo {E.~D'Hoker and D.H.~Phong, {\it The geometry of string
perturbation theory }, \rmp60(1988)917.}
\lr\bfkov {C.~Bachas, C.~Fabre, E.~Kiritsis, N.~Obers, P.~Vanhove, {\it
    Threshold effects and D-instantons} to be published.}
\lr\jacob{Jacobson, {\it Basic Algebra}, Vol. I.}
\lr\witten{E.~Witten, {\it String theory in various dimensions},
\xxx9503124,
  \npb443(1995)85.}
\lr\luc{M.~L{\"u}scher, K.~Symanzik and P. Weisz, {\it Anomalies of the
free loop wave equation in the WKB approximation}, \npb173(1980)365.}
\lr\polI{ J.~Polchinski, {\it Evaluation of one loop string path
integral},
\cmp104(1986)37.}
\lr\weis{ W.~Weisberger, {\it Normalization of the path integral measure
and the coupling constants for bosonic strings}, \npb284(1987)171.}
\lr\wein{ S.~Weinberg, {\it Coupling constants and vertex functions in
string theories}, \plb156(1985)309.}
\lr\witt{E.~Witten, {\it Bound States Of Strings And $p$-Branes},
\xxx9510135,
\npb460(1996)335.}
\lr\julia{ E. Cremmer, B. Julia and J. Scherk, {\it Supergravity
theory in
eleven dimensions}, Phys. Lett. {\bf 76B} (1978) 409.}
\lr\gw{D.J.~Gross and E.~Witten, {\it Superstring modifications of
Einstein's
    equations}, \npb277(1986)1.}
\lr\gris{M.T.~Grisaru , A.E.M~Van de Ven and D.~Zanon, {\it
Two-dimensional
supersymmetric sigma models on Ricci flat Kahler manifolds are not
finite},
\npb277(1986)388 ; {\it Four loop divergences for the N=1 supersymmetric
nonlinear sigma model in two-dimensions}, \npb277(1986)409.}
\lr\maclean{B.~McClain and B.D.B.~Roth, {\it Modular invariance for
    interacting bosonic strings at finite temperature},
\cmp111(1987)539.}
\lr\zag{D.~Zagier, {\it Eisenstein series and the Riemann zeta function},
in Tata Inst. Fund. Res. Studies in Math., 10, Tata Inst.
Fundamental Res.,
Bombay, 1981.}
\lr\unit{N.~Sakai and Y.~Tanii, {\it One-loop amplitudes and
effective action
in
superstring theories}, \npb287(1987)457.}
\lr\polchtasi {J.~Polchinski, {\it TASI Lectures on D-branes},
\xxx9611050.}
\lr\greenhove{M.B. Green and P. Vanhove, in preparation.}
\lr\terras{A.~Terras, {\it Harmonic Analysis on Symmetric Spaces and
    Applications}, vol.~I, Springer--Verlag 1985.}
\lr\vafawitten{C.~Vafa and E.~Witten, {\it A one loop test of
string duality},
  \xxx9505053, \npb447(1995)261.}
\lr\duffminas{M.J.~Duff, J.T.~Liu and R.~Minasian, {\it
Eleven-dimensional
    origin of string-string duality: a one loop test}, \xxx9506126,
  \npb452(1995)261.}
\lr\douglasetal{M.R.~Douglas, D.~Kabat, P.~Pouliot and S.~Shenker, {\it
D-branes and short distances in string theory}, \xxx9608024,
\npb485(1997)85.}
\lr\kabat{D.~Kabat and P.~Pouliot {\it A comment on
zero-brane quantum
mechanics}, \xxx9603127, \prl77(1996)1004.}
\lr\kiritsis{E.~Kiritsis
Private communication}
\lr\swedes {U.H.~Danielsson, G.~Ferretti and
B.~Sundborg, {\it D-particle
Dynamics and Bound States}, \xxx9603081, \ijmp11(1996)5463.}
\lr\horavawit{P. Horava and E. Witten, {Eleven-dimensional
supergravity on
a manifold with boundary},  Nucl. Phys. {\bf B475} (1996) 94,
\xxx9603142.}
\lr\lerche{W.  Lerche, {\it Elliptic index and superstring
effective actions},
Nucl. Phys. {\bf B308} (1988) 102.}
\lr\tseytlin{A.A. Tseytlin, {\it Heterotic type I superstring duality
and low-energy effective actions}, Nucl. Phys. {\bf B467}
(1996) 383, \xxx9512081.}
\lr\roo{M. de Roo, H. Suelmann and A. Wiedeman, {\it Supersymmetric $R^4$
actions in ten-dimensions}, Phys. Lett. {\bf B280} (1992) 39.}
\lr\tthree{D. Berenstein, R. Corrado and J. Distler, {\it On the moduli
spaces of M(atrix)-theory compactifications}, \xxx9704087.}

\noblackbox
\baselineskip 14pt plus 2pt minus 2pt
\Title{\vbox{\baselineskip12pt
\hbox{hep-th/9704145}
\hbox{DAMTP/97-31}
\hbox{CPTH-S501-0497}
}}
{\vbox{
\centerline{D-INSTANTONS, STRINGS AND M-THEORY} }}

\centerline{ Michael B. Green,}
\medskip
\centerline{DAMTP, Silver Street, Cambridge CB3 9EW, UK} \centerline{\it
M.B.Green@damtp.cam.ac.uk} \bigskip
\centerline{Pierre Vanhove}
\medskip
\centerline{Centre de Physique Th{\'e}orique, } \centerline{Ecole
Polytechnique, 91128 Palaiseau, FRANCE} \centerline{\it
vanhove@cpth.polytechnique.fr}

\medskip
\centerline{{\bf Abstract}}

The $R^4$ terms in the effective action for M-theory compactified on
a two-torus are motivated by combining one-loop results in type II
superstring
theories with the $\Sl2$ duality symmetry. The conjectured expression
reproduces precisely the tree-level and one-loop $R^4$ terms in the
effective action of the type II string theories compactified on a circle,
together with the expected infinite sum of instanton corrections. This
conjecture
implies that the $R^4$ terms in
ten-dimensional string type II theories receive no perturbative
corrections
beyond one
loop and there are also no non-perturbative corrections in the
ten-dimensional IIA
theory. Furthermore, the eleven-dimensional M-theory limit exists,
in which
there is an $R^4$ term that originates entirely from the one-loop
contribution
in the
type IIA theory and is related by supersymmetry to the eleven-form
$C^{(3)}R^4$.
The generalization to compactification on $T^3$ as well as
implications for
non-renormalization theorems in D-string and D-particle interactions are
briefly discussed.

\noblackbox
\baselineskip 14pt plus 2pt minus 2pt

\Date{1997}

\newsec{Introduction}

The interconnections between apparently distinct superstring
theories and their
connection to eleven-dimensional M-theory provide strong constraints on
their non-perturbative structure. Any of the various string `theories' is
defined as a perturbative expansion in powers of the string coupling,
$e^\phi$, where $\phi$ is the dilaton. Its low energy behaviour is
determined by an effective action that is a function of background
massless
fields obtained by integrating out all (massless and massive) quantum
fields. The effective action has an expansion in powers of space-time
derivatives (inverse powers of the string tension). The absence of
a scalar
field in
eleven-dimensional M-theory
means that it does not possess a loop expansion but it does have
an effective action that may, in principle, be expressed as a low energy
expansion that begins with the standard supergravity action of
\refs{\julia},
which is a supersymmetric extension of the Einstein--Hilbert action,
\eqn\einhil{S_R = {1\over 2\kappa_{11}^2} \int d^{11} x \sqrt{-G^{(11)}
}R,} where $G^{(11)}$ is the determinant of the eleven-dimensional
metric.\foot{In the following the ten-dimensional string
metric will
be denoted by lower-case $g$ while the M-theory metric will be denoted by
an upper-case $G$. In both cases our convention is that the metric is dimensionless.}

It was argued in \refs{\witten} that this may be viewed as the strong
coupling limit of
ten-dimensional type IIA superstring theory with the identification
\eqn\witrel{R_{11} = (\al)^{1/2}\lambda^{A},} where
$\lambda^A$ is
the type IIA coupling constant and $R_{11}$ is the radius of the eleventh
dimension in the M-theory metric.

The next gravitational terms in the low-energy expansion of type II
superstring actions
beyond the Einstein--Hilbert term are fourth order in the Riemann
curvature.  In this paper we shall use the symbol $\Rfour$ to  indicate
these terms, in which the contractions of the four Riemann tensors
are defined
by
\eqn\tensterm{\Rfour\equiv t^{\mu_1\dots \mu_8} t_{\nu_1\dots \nu_8}
R^{\nu_1\nu_2}_{\mu_1\mu_2} \cdots R^{\nu_7\nu_8}_{\mu_7\mu_8},}
and the
tensor $t^{\mu_1\dots \mu_8}$ ($\mu_r = 0,1,\cdots,9$) will be defined
later.
  These are the leading terms in the low energy
limit of one-loop amplitudes \refs{\gs} and the first non-leading
corrections to the
low-energy limit at tree level \refs{\gw,\gris}.    Furthermore, it was
shown in  \refs{\gg}\ that terms of this form are induced in
ten-dimensional type IIB  superstring theory by integration over the
fermionic zero modes in a D-instanton background.  The arguments of this
paper will strongly suggest that the exact form of the $R^4$ terms in the
effective nine-dimensional action
for M-theory compactified on $\calt^2$ is
\eqn\mainres{S_{R^4}
={1\over 3\cdot (4\pi)^7 l_{11}}  \int d^9 x
\sqrt{-G^{(9)}}\left(  \C|V_2^{-1/2} f(\Omega,
\bar \Omega) + {2\pi^2\over 3} \C|V_2\right) \Rfour,}
where $G^{(9)}_{mn}$ is the nine-dimensional metric, $\Omega$ is
the complex structure of $\calt^2$ and $4\pi^2l_{11}^2 \C|V_2$ is its volume.
The modular function $f(\Omega, \bar \Omega)$ is the same non-holomorphic
Eisenstein series as the one that was conjectured in \refs{\gg} to determine
the $R^4$ term in ten-dimensional type IIB superstring theory, with $\Omega$
replaced by 
\eqn\rhobdef{\rho^B \equiv \rho_1^B + i \rho_2^B = C^{(0)} + i
e^{-\phi^B},}
where $C^{(0)}$ is the \RR\ pseudoscalar and $\phi^B$ the IIB dilaton.
In equation~\mainres\ the eleven-dimensional Planck length, $l_{11}$ is related
to $\kappa_{11}$ by $\kappa_{11}^2 = (2\pi)^8 l_{11}^9 /2$.  We will later
use the normalization of the ten-dimensional string theory coupling in
which
\eqn\kdef{\kappa_{10}^2 = {\kappa_{11}^2\over2\pi R_{11} (\lambda^A)^2} = 2^6 \pi^7
\alpha^{\prime
4} ,}
which makes the tension of the D-string equal to $e^{-\phi} \times$
(tension of the fundamental string)  \refs{\polchtasi}.   From  now
on we will
set $\alpha'=1$ to simplify the expressions.

The expression \mainres\ will reinforce the conjecture in \refs{\gg}
that the
ten-dimensional IIB
theory satisfies a perturbative non-renormalization theorem ---
there are no
contributions beyond one loop and the non-perturbative contributions are
determined
by multiply-charged D-instantons. The $\Sl2$ duality symmetry of the IIB
theory will be related to the geometry of the torus as in
\refs{\aspinwall,\schwarz}.
All non-perturbative effects will be seen to disappear in the
ten-dimensional
type IIA theory, essentially because there are no finite-action
instantons,
and the $R^4$ term is then given entirely by the sum of the perturbative
tree-level and one-loop term. The decompactified
eleven-dimensional M-theory effective action ($\C|V_2 \to \infty$) has an
$R^4$ term that comes
entirely
from the one-loop type IIA term with a coefficient that is fixed
precisely by
the one-loop string calculation. The complete effective action for
M-theory
could then be
determined, in principle, by eleven-dimensional supersymmetry which
should
relate the new term to the Einstein--Hilbert term.

\newsec{$R^4$ terms in type II superstring perturbation theory.}
Consideration of the on-shell scattering of four gravitons in either type
IIA or IIB string perturbation theory at tree level
\refs{\gris,\witten} and
one loop \refs{\gs} leads to terms in the low energy Lagrangian of the
form\foot{
Expressions in which the superscript/subscript $A$ or $B$ is
omitted applies to
either type II theory}
$\sqrt {-g} \Rfour$ which are $O(\alpha^{\prime -1})$ whereas the leading
term is the
Einstein--Hilbert action given by
\eqn\einhilstring{{1\over 2\kappa_{10}^2} \int d^{10}
xe^{-2\phi}\sqrt {-g}
R.}

After compactification on an $n$-torus, $\calt^n$, the sum of the
tree-level and one-loop contributions to the four-graviton amplitude has
the form
\refs{\greenbrink}
\eqn\pertl{
A = K_{in} {\kappa_{10}^2 \over 12\cdot
2^8}\left[-\lambda^{-2}\C|V_n{\G(- s/4)\G(- t/4)\G(- u/4)\over
\G(1 + s/4)\G(1+ t/4)\G(1+ u/4)} + {\kappa_{10}^2\over 2^5
\pi^6 } d_1 \right] }
where the coupling $\lambda$, is determined by the expectation value,
$\langle\phi \rangle$, of the dilaton,
\eqn\lambdadef{\lambda =e^{ \langle\phi \rangle},}
and $\C|V_2 =\sqrt{- g^{(n)}}$  is the volume of the
compactified space
with metric $g^{(n)}_{ij}$.
The relative normalization between the tree level and the one-loop
term in
\pertl\ was
determined by arguments based on unitarity in \refs{\unit}.

The kinematic factor in \pertl\ is eighth order in the momenta and can be
written as
\eqn\kin{
K_{in} \sim \hat t_R^{\mu_1 \mu_2 \cdots \mu_8} \hat
t_S^{\nu_1\nu_2 \cdots \nu_8}
\zeta^{(1)}_{\mu_1 \nu_1} k^{(1)}_{\mu_2} k^{(1)}_{\nu_2} \cdots
\zeta^{(4)}_{\mu_7 \nu_7} k^{(4)}_{\mu_8} k^{(4)}_{\nu_8}, } where
$k^{(r)}_{\mu}$ is the momentum of the graviton labelled $r$ and
$\zeta_{\mu_r \nu_r}^{(r)}$ is its polarization tensor. The constant
$\kappa_{10}$ depends on the definition of the dilaton. The eighth-rank
tensors $\hat t_R$ ($R=1,2$) are conveniently defined in a
light-cone frame
by
\eqn\tdefs{\eqalign{ \epsilon_{a_1a_2\cdots
a_8}\gamma^{i_1j_1}_{a_1a_2}\cdots
\gamma^{i_4j_4}_{a_7a_8} & =\hat t_1^{i_1j_1\cdots i_4j_4}=
t^{i_1j_1\cdots
i_4j_4}+{1\over2}\epsilon^{i_1j_1\cdots
j_4j_4} \cr
\epsilon_{\dot a_1 \dot a_2\cdots \dot a_8}\gamma^{i_1j_1}_{\dot a_1\dot
a_2}\cdots \gamma^{i_4j_4}_{\dot a_7\dot a_8} & =\hat t_2^{i_1j_1\cdots
i_4j_4}= t^{i_1j_1\cdots i_4j_4} - {1\over2}\epsilon^{i_1j_1\cdots
j_4j_4}
,\cr}}
where $a$ and $\dot a$ are $SO(8)$ indices labelling the ${\bf 8}_s$ and
${\bf 8}_c$ representations and $i_r,j_r = 1, \cdots, 8$ label the ${\bf
8}_v$ representation. The vector indices are covariantized in the
ten-dimensional expression, \kin. In the type IIB theory the two
$\hat t_8$'s
are the same ($R=S$) in \kin, leading to an irrelevant ambiguity in the
sign of the $\epsilon_8 \epsilon_8$ term,  whereas they are
different ($R\ne
S$)  in
the IIA theory. The $\epsilon^{i_1j_1\cdots j_4j_4}$ terms are total
derivatives and we will discard them in the following, in which
case there
is no perturbative distinction between the $R^4$ terms in the two
theories.
Since, at linearized level, $R^{\mu\nu}_{\rho\sigma}=\kappa_{10}
k^{[\mu}\zeta^{\nu]}k_{[\rho}\zeta_{\sigma]}$ the kinematical factor is
$K_{in}=t_8t_8R^4/(24\ \kappa_{10}^2)$.

The coefficient, $d_1$, of the one-loop term in \pertl\ is given by the
integral of a
modular function over the fundamental domain of $\Sl2$. When the
theory is
compactified on an $n$-torus it takes the form, \eqn\Loop{
d_1= \int_\C|F {d^2\tau\over \tau_2^2} Z_{lat} F(\tau,\bar \tau)} where
$Z_{lat}$ is the partition function associated with the lattice,
$\Gamma^{n,n}$,
\eqn\latII{
Z_{lat} = \C|V_n \sum_{m,n \in \ZZ} e^{-{\pi\over\tau_2}\sum_{i,j}
(g+B)_{ij}
(m_i + n_i \tau) (m_j + n_j
\bar\tau)}
}
and $i,j = 1,\dots, n$ label the directions in the lattice. This
sum can be
interpreted as the
sum of contributions to the functional integral from fundamental string
world-sheets in which the two world-sheet coordinates wind $m_i$
and $n_i$
times around the compact dimension.
The dynamical factor in \Loop\ is given by \eqn\fdef{
F(\tau,\bar \tau) = {1\over \tau_2^3} \int_{\calt^2} \prod_{i=1}^3
d^2\nu_i
\left[\chi_{12}\chi_{34} \over\chi_{13}\chi_{24} \right]^{-
s}\left[\chi_{14}\chi_{23} \over\chi_{13}\chi_{24} \right]^{- t}, }
where $\ln \chi_{ij}$ is the scalar Green function between the vertices
labelled $i$ and $j$ on the toroidal world-sheet, $\calt^2$ (and
$\int_{T^2} \prod d^2 \nu_i = \tau_2^3$).

The leading low energy contributions obtained by expanding \pertl\ at in
powers of momenta, are the massless pole terms and the contact term that
are associated with the (linearized) Einstein--Hilbert action. After
subtracting these terms
the remainder of \pertl\ gives the $R^4$ terms which are obtained by
setting the momenta to zero inside the square brackets. The tree
contribution to the effective action is,
\eqn\treer{S_{R^4}^{tree} = {\zeta(3)\over 3\cdot 2^7\kappa_{10}^2}\C|V_n
\int
d^{10-n} x \sqrt {- g^{(10-n)}} (\rho_2)^2 \Rfour,} in a normalization
consistent with \einhilstring. We have replaced $\lambda$ by
$\rho_2^{-1} =
e^\phi$ in this expression as we will do in the following.
The loop contribution (the toroidal world-sheet) in \pertl\ depends on
details of the the compactification, which will now be considered.

\newsec{Compactification to nine dimensions on $S^1$.}

The one-loop term in the nine-dimensional theory is obtained by setting
$g_{10\, 10} =
r^2$ and $B=0$ in \latII, where $r$ is the circumference of the tenth
dimension in sigma model (string frame) units so that $\C|V_1= r$. After
performing a
Poisson resummation on one integer \Loop\ gives \eqn\looptwo{
d_1= r \int_\C|F {d^2\tau\over\tau_2^2}
\sum_{(m,n)\in\ZZ^2} e^{- \pi r^2 |m\tau +n|^2/\tau_2} F(\tau,\bar
\tau). }

Following a standard procedure \refs{\maclean,\dkl} it is useful to
separate
the term with $(m,n)= (0,0)$ and set $m = sp$ and $n =s q$ in the other
terms where $p$ and $q$ are coprime integers and $s$ is an unconstrained
integer. The sum over $p,q$ is a sum over fundamental domains of
$\Sl2$ which
is equivalent to extending $\C|F$ to the semi-infinite strip, $0
\le \tau_2
\le \infty$, $-1/2 \le \tau_1 \le 1/2$, so that \looptwo\ can
be expressed as \refs{\bfkov},
\eqn\loopthree{
d_1= r \left[\int_\C|F {d^2\tau\over
\tau_2^2}F(\tau)
+\int_{Strip} {d^2\tau\over\tau_2^2}\sum_{s\in\ZZ\backslash\{0\}}
e^{- \pi
r^2 s^2/ \tau_2} F(\tau)\right].
}

The integrals converge and using the low energy limit in \fdef,
which sets
$F=1$, the result of the integrations is \eqn\gonedef{
d_1= {\pi \over 3} \left[ r + {1 \over r } \right].
}The total contribution of the $R^4$ terms to the effective action
therefore
has the form

\eqn\genact{
\C|S_{R^4} ={1\over 3\cdot 2^8 \kappa_{10}^2} \int d^9 x \sqrt {-g^{(9)}}
\, \Rfour\, r \left[2\zeta(3)
(\rho _2)^2+ {2 \pi^2\over 3 } ( 1 + {1\over r ^2} ) + \cdots \right], }
where $\cdots$ represents potential higher-order perturbative and
non-perturbative terms.

The same expressions apply to the type IIA and the type IIB
theories which are
related by the T-duality transformations, \eqn\tdual{ r_A = {1\over
r_B }, \qquad r_Ae^{-\phi_A} = e^{-\phi_B}, \qquad C^{(1)} = C^{(0)}, }
where the subscript $A$ or $B$ indicates which theory the relevant
quantities
are defined in and $C^{(1)}\equiv C^{(1)}_{10}$ is the component of
the IIA
\RR\ vector potential in the tenth direction. The value of
$C^{(0)}$ in the
IIB theory does not enter into the fundamental string amplitudes
but it is
related to the component of the \RR\ vector of IIA via \tdual\ and
hence to
the complex structure of the torus in the compactification of
M-theory to nine
dimensions on $\calt^2$ to be described later. The $\Sl2$ symmetry
of type IIB
implies that the effective action is invariant under integer shifts of
$C^{(0)}$ which implies that the IIA action must be invariant under
$C^{(1)}
\to C^{(1)} + 1$.       The complex scalar,
\eqn\rhoadef{\rho^A \equiv \rho^{A}_1 + i \rho^A_2 = C^{(1)} + i {r_A}
e^{-\phi_A}, }
is equated with $\rho^B$ by the T-duality transformation, \eqn\relrho{
\rho^A = \rho^B.}

We must also consider non-perturbative contributions to the
$t_8t_8R^4$ term
due to the effects of D-instantons. The type II theories have a
total of 32
components in their supercharges. Since a D-instanton breaks half of the
supersymmetries, there are at least sixteen fermionic zero modes in the
fluctuations around instanton configurations \refs{\bfkov,\ggII}.
However, the
$t_8t_8R^4$ term arises only from the sector with sixteen fermionic
zero modes
\refs{\gg}. Therefore, we only need to consider configurations with
single
D-instantons carrying multiple charges.
This will turn out to be consistent with the various duality
symmetries in the
problem and and with the coefficients of the perturbative terms in
\genact.
 From the point of view of the IIA theory in nine dimensions the only
instantons are configurations in which the Euclidean world-line of a
ten-dimensional $D0$-brane winds around the tenth dimension \refs{\gg}.
Since the $D0$-branes are Kaluza--Klein modes of M-theory there must be a
single normalizable $D0$-brane state with charge $n$ and mass
proportional
to $n$ (this is the basis of the as yet unproven conjecture
that there is precisely one threshold bound state of $n$ minimally
charged
$D0$-branes \refs{\witt}). In the Euclidean compactification to
nine dimensions
the world-line of such a particle can wind $m$ times so that its
action is
$2\pi mn \rho^A$. The consequent non-perturbative terms in the effective
action have the form \refs{\gg},
\eqn\sumnona{\sum_{m,n > 0} c^A_{mn}(\rho_2^A, r_A) \left(e^{2\pi i
mn \rho^A}
+ e^{-2\pi i mn \bar\rho^A}\right),}
which is consistent with the shift symmetry of $C^{(1)}$ and the
coefficients
$c^A_{mn}$ are to be determined. These coefficients can be determined
directly by evaluating the functional integral for a supersymmetric
D-particle world-line that is wrapped around the compact Euclidean
direction. Further details will
be given in \refs{\greenhove} but here we will determine the coefficients
$c_{mn}^A$ by a duality argument.

T-duality equates the series \sumnona\ with the series of D-instanton
contributions to the type IIB theory,
\eqn\sumnonb{\sum_{m,n > 0} c^B_{mn}(\rho_2^B, r_B) \left(e^{2\pi i
mn \rho^B}
+ e^{-2\pi i mn \bar\rho^B}\right),}
which was discussed in \refs{\gg} in the $r_B\to \infty $ limit. In that
limit the complete $R^4$ term of the ten-dimensional IIB effective
action has
the form,
\eqn\rfours{ S_{R^4} = {1\over 3\cdot 2^8 \kappa_{10}^2} \int d^{10} x
\sqrt{-g^B} (\rho_2^B)^{1/2} f(\rho^B,\bar \rho^B) \Rfour,}
where the function $f(\rho^B,\bar \rho^B) $ must be modular
invariant since
the group $\Sl2$ is a duality symmetry of
type IIB superstrings in the Einstein frame \refs{\hulltownsend} under
which \eqn\rhotrans{\rho^B \to {a \rho^B + b\over c\rho^B + d},} where
$ad-bc=1$ and the coefficients are integers (and $R$ is inert). We
are here
using the fact that
\eqn\einsdef{\sqrt{-g^B_E}\, \Rfour = \sqrt{-g^B} (\rho^B_2)^{1/2}\,
\Rfour,} where $g^B_E$ is the Einstein-frame metric.

A conjecture was
made in
\refs{\gg} that $f(\rho^B, \bar \rho^B) =\zeta(3) E_{3\over 2}(\rho^B)$,
where $\zeta$ is the Riemann zeta function and $E_s(\rho)$ is a
non-holomorphic Eisenstein series (or Maass waveform) defined by
(\refs{\zag}, \refs{\terras})
\eqn\edefs{E_s(\rho) = \sum_{\gamma\in \Gamma /\Gamma_\infty}
\left[\Im (
\gamma \rho) \right]^s ,} where $\gamma$ indicates a transformation in
$\Gamma = \Sl2$ modded out by the subgroup defined by  $\Gamma_\infty =
\pmatrix{\pm 1 & n\cr 0 & \pm 1}$. Such Eisenstein series are
eigenfunctions of the Laplace operator on the fundamental domain of
$\Sl2$,
\eqn\laplaceq{\Delta E_s(\rho) \equiv \rho_2^2\left({\partial^2 \over
\partial \rho_1^2} + {\partial^2 \over \partial \rho_2^2} \right)
E_s(\rho)
= s(s-1) E_s(\rho).} The function $f$ can be
expressed in
various ways as
\eqn\bessum{\eqalign{ & f (\rho^B, \bar \rho^B) = \zeta(3)
\sum_{\gamma\in \Gamma /\Gamma_\infty} \left[\Im ( \gamma \rho^B)
\right]^{3/2} = \sum_{(p,n )\neq
(0,0)}{(\rho^B_2)^{3/2}\over |p+n\rho^B|^3} \cr &\ \quad\qquad\quad =
2\zeta(3)(\rho^{B }_2)^{3/2} + {2\pi^2\over 3}(\rho^B_2)^{-1/2}
+ 8\pi (\rho_2^B)^{1/2}\sum_{m\not=0,n\geq 1} e^{2i\pi nm \rho_1}
\left|{m\over n}\right|
K_1\left(2\pi |m|n \rho_2^B \right) \cr
&\ \quad\qquad\quad = 2\zeta(3)(\rho^{B }_2)^{3/2} + {2\pi^2\over
3}(\rho^B_2)^{-1/2}
\cr
& + 4\pi \sum_{m,n \ge 1} \left({m\over n^3}\right)^{1/2}
\left(e^{2\pi i mn
\rho^B} + e^{-2\pi i mn \bar \rho^B} \right) \left(1 + \sum_{k=1}^\infty
(4\pi mn
\rho^B_2)^{-k} {\Gamma( k -1/2)\over \Gamma(- k -1/2) k!} \right) , \cr}}
where $K_1$ is a Bessel function. Intriguingly, the first
expression on the
right-hand side of \bessum\ has the form of the tree-level term of
\refs{\gw,\gris} summed over all its $\Sl2$ images -- in other words, $
( \rho_2^B)^{3\over 2}\sum_{p,q} T^{-3}_{pq}$, where $p,q$ are coprime
and $T_{pq}$ is the tension in the $(p,q)$ dyonic string. The first two
terms in the last expression for $f$ in \bessum\ should be compared with
the ten-dimensional perturbative tree-level term and one-loop terms in
\pertl. The conjectured agreement requires the precise relative
normalization of the tree-level and one-loop terms \refs{\unit}.
The remaining infinite series represents the sum over a dilute gas of
multiply-charged D-instantons and anti D-instantons that converges for
large $\rho^B_2$ (small coupling). Various motivations for this
expression were
described in \refs{\gg}.

We will here generalize this description and obtain more insight by
making
use of the fact that in the
nine-dimensional theory
the $\Sl2$ symmetry of the IIB string theory can be interpreted as a
geometric symmetry of M-theory compactified on a torus
\refs{\aspinwall,\schwarz}. To see this it is necessary to translate the
coordinates to the M-theory frame -- the frame in which
eleven-dimensional
supergravity is naturally formulated. Following \refs{\witten} the
eleven-dimensional metric may be parameterized by \eqn\eleveng{ ds^2=
G^{(10)}_{mn} dx^mdx^n + R_{11}^2 (dx^{11}- C^{(1)}_m dx^m)^2,}
where the ten-dimensional part of the metric is $G^{(10)}_{mn}=
R_{11}^{-1} g^A_{mn}$ (recalling that $g^A_{mn}$ is the
ten-dimensional IIA
metric in the
string frame). Compactifying this on a circle of radius $R_{10}$
leads to the
equivalences,
\eqn\rad{\eqalign{
g^A_{10\, 10} = r_A^2 &=  R^2_{10}R_{11}=
G_{10\, 10} R_{11}, \qquad  \rho_2^A
= R_{11}^{-3/2} \cr &  \rho_2^B ={R_{10}\over
R_{11}}, \qquad r_B = { 1\over R_{10} \sqrt {R_{11}}}.\cr }}

Using a block diagonal ansatz for the eleven-dimensional metric it can be
written so that
\eqn\met{\sqrt{-G^{(11)}} = \sqrt{G^T} \sqrt {- G^{(9)}} =l_{11}^{-2} R_{10} R_{11}
\sqrt{- G^{(9)}} = \C|V_2 \sqrt{-G^{(9)}}, } where
$4\pi^2l_{11}^2\C|V_2 =4\pi^2 R_{10}R_{11}$ is the volume of
$\calt^2$. The metric on the two-torus, 
\eqn\mtorus{ G^T ={1\over l_{11}^2} \pmatrix{ R_{10}^2 + R_{11}^2(C^{(1)})^2 & -R_{11}^2
C^{(1)} \cr -R_{11}^2 C^{(1)} & R_{11}^2 \cr}} can be expressed in
terms of
string frame quantities so its complex
structure is given by
\eqn\mcomp{\eqalign{\Omega = \Omega_1 + i \Omega_2 & = C^{(1)} + i
{R_{10}
\over R_{11}} \cr
& = C^{(1)} + i {r_A} e^{-\phi^A} = \rho_A\cr
& = C^{(0)} + i
e^{-\phi^B} =\rho_B .\cr}}

Equation \genact\ can now be rewritten in coordinates appropriate for the
type IIA, IIB and M-theory as,\foot{Recall that we are setting $\alpha'=1$
  in~\kdef.}
\eqn\mact{\eqalign{
\C|S_{R^4} &= {1\over 3\cdot 2^8 \kappa_{10}^2} \int d^{9} x \sqrt{-
g^{A(9)}}\, \Rfour\, r_A\, \left[ 2 \zeta(3) (\rho^A_2)^2 + {2\pi^2
\over 3
}(1 + {1\over r_A^2}) + \cdots \right] \cr
&= {1\over 3\cdot 2^8 \kappa_{10}^2} \int d^{9} x \sqrt{- g^{B(9)}}\,
\Rfour\, r_B\, \left[2\zeta(3) (\rho^B_2)^2 + {2\pi^2\over 3 } (1 +
{1\over r_B^2}) + \cdots \right]
\cr
&= {l_{11}^6 \over 3\cdot  2^8\kappa_{11}^2}  \int d^9 x \sqrt{- G^{(9)}}\, \Rfour\,
2\pi R_{11} R_{10}
\left[2\zeta(3){l_{11}^3\over R_{11}^3}+ {2\pi^2\over 3 } +
{2\pi^2\over 3}{l_{11}^3\over R^2_{10}R_{11}} + \cdots \right] \cr}}
where $ g^{A(9)}$, $ g^{B(9)}$ are the nine-dimensional metrics in
the IIA
and IIB theories and the non-perturbative terms, represented by
$\cdots$, are
given by a power series
in $e^{2\pi i \rho^A}$, $e^{2\pi i \rho^B}$ and $e^{2\pi i \Omega}$,
respectively.

Since we know that the last expression in \mact\ must be invariant
under the
action of $\Sl2$ on $\Omega$ it is appealing to write it as an
expansion for
large $\Omega_2$,
\eqn\magain{\eqalign{\C|S_{R^4} = {1\over 3\cdot (4\pi)^7 \, l_{11}}
    \int d^9 x & \sqrt{- G^{(9)}} \, 
\Rfour\cr &
\left\{  \C|V_2 ^{-1/2} \left[2\zeta(3) (\Omega_2
)^{3/2} + {2\pi^2 \over 3 } (\Omega_2 )^{ -1/2} +
\cdots \right] + {2\pi^2 \over 3 } \C|V_2 \right\}, \cr} }
which should be identified with the expansion of a modular function of
$\Omega$.
We now compare this with the expansion of the modular function
in \mainres\ where the function $f$ is defined in \bessum.  We see
that the
perturbative terms in \magain\  are
identical to those in \mainres.
Correspondingly, non-perturbative extensions of the effective $R^4$
actions in
the IIA and IIB theories follow by substituting the appropriate
variables in \mainres.
Several striking features are apparent from the structure of \mainres:

$\bullet$ There are only two perturbative terms in the expansion of $f$,
corresponding to the tree and the one-loop terms in the fundamental
string
calculations. This points to a perturbative non-renormalization theorem
beyond one loop in type II string theory. It would be
gratifying to demonstrate this explicitly from the expressions for
superstring
perturbation theory at higher genus but this seems to be
difficult.\foot{We are
grateful to Nathan Berkovitz for correspondence on this issue.} Such a
non-renormalization theorem can be motivated heuristically as follows.
The
four gravitons attached to a torus are just sufficient to soak up the
sixteen zero
modes of the space-time fermions. At higher genus there could
easily be some
extra fermionic zero modes leading to a vanishing result for the
effective
$\Rfour$ term in the low energy ($\al \to 0$) limit. The complication is
that, at least in the
light-cone formalism, there are vertex insertions that might soak these
surplus zero
modes up.

$\bullet$ In the limit $r_B\to \infty$ \mainres\ reduces to the sum over
non-perturbative terms (in the string frame) conjectured in \refs{\gg}
based on the
properties of D-instantons in the ten-dimensional type IIB theory.

$\bullet$ In the limit $r_A \to \infty$ \mainres\ reduces to the
first two
terms in square parentheses in the first expression in \mact. All the
non-perturbative contributions vanish and so the full expression for the
ten-dimensional type IIA theory has just the perturbative tree-level and
one-loop terms.

$\bullet$ Upon decompactifying the M-theory torus, taking $\C|V_2
\to \infty$
in \mainres, only the term proportional to $\C|V_2$ contributes -- the
constant term in the square parentheses in \mact. The result is that the
$R^4$ term in the M-theory effective action is determined precisely
by the
coefficient of the one-loop diagram in the type IIA superstring
theory and is
given by
\eqn\mresult{S_{R^4} = {1\over 18\cdot (4\pi)^7\, l_{11}^3}
\int d^{11}x \sqrt{-G^{(11)}} \Rfour}
in a normalization in which the Einstein--Hilbert action is given
by \einhil.

 The range of the indices in $t_8$ is here extended trivially in  
the eleventh dimension so that $t_8 \equiv t^{\mu_1 \cdots \mu_8}$  
with $\mu_r = 0, \cdots, 10$.

Since \mainres\  has a finite
M-theory limit as $\C|V_2 \to \infty$ and reduces to the correct tree and
one-loop terms for the IIA and IIB theories as $r_A$ or $r_B \to \infty$
and also satisfies the correct T-duality relation between  the type
IIA and
type IIB theories, it is a good candidate for the exact $R^4$ term.   We
should, however, consider to what extent these conditions determine the
solution
uniquely.

In principle, we could add a function $h (\C|V_2;\Omega,\bar
\Omega)$ to the
 terms in parentheses in \mainres, which must be a modular function of
 $\Omega$ and must not spoil the above properties.   The existence of the
 M-theory  limit means that $h \sim (\C|V_2)^\alpha k(\Omega,\bar
\Omega)$
 (where $k$ is a modular function) as $\C|V_2 \to \infty$ with
$\alpha < 1$.
 However, if this limit can be
interchanged with the perturbative type IIA limit  then  $h \sim
(\C|V_2)^\alpha \Omega_2^\beta \sim (r_A)^{\alpha + \beta}
(\rho^A_2)^{\beta - \alpha/3}$ as $\rho^A_2 \to \infty$.  Taking into
account a power of $(R_{11})^{-1/2}$ from the measure in \mainres\
to go to
 the string frame,
the net power of $\rho^A_2$ is   $X = \beta -(\alpha - 1)/3$.
However, this  only contributes  in ten dimensions  if $\alpha +
\beta =1$,
so that $X >0$  which  spoils the known perturbative behaviour.    This
excludes the presence of terms beyond the tree and one-loop terms in the
ten-dimensional IIA theory subject to  the (very strong) assumption
of the
uniformity of the M-theory and perturbative limits.

 More generally, the function $h$ may  vanish  in the M-theory limit
($\alpha < 1$) and have mild enough perturbative behaviour not to
spoil the tree or  one-loop terms.  In that case it is an $L^2$
function on
the fundamental domain of $\Sl2$ and, following  \refs{\terras},
it can be
written as a sum of cusp forms and a continuous integral over $E_s$.
Superficially, the constraints imposed by T-duality and by the
consistency
of the various limits do not exclude the addition of $h$ to
\mainres.
They would be eliminated if we had a reason to require $f$ to satisfy the
eigenvalue equation, \laplaceq, which has  $f$ as its unique
solution for a
given eigenvalue (as follows from theorem 1
section 3.5 of \refs{\terras}). Although this equation
has not been motivated by a direct argument, it has the flavour of a
condition that might follow by requiring supersymmetry
of the effective action.

Having described the nine-dimensional theory in detail it is of
interest to
understand the extended U dualities of theories obtained by
compactification to
lower dimensions. These provide further consistency checks on the
validity
of the nine-dimensional expression.

\newsec{M-theory on $\calt^3$ or IIB on $\calt^2$.}

Upon compactifying to eight dimensions there is a richer spectrum of
instantons. In addition to the direct reduction of the instantons
from nine
dimensions
there are those that arise from the M-theory/IIA point of view from the
wrapping
of the Euclidean three-volume of the $M2$-brane and from the IIB
side from
the wrapping of the $\Sl2$ multiplet of Euclidean world-sheets of the
fundamental and D-strings.
The structure of the duality group, $\Sl3 \times \Sl2$, is
correspondingly
richer in eight dimensions.

We will concentrate on the compactification of the IIB theory on ${\cal
T}^2$ but to begin with we will briefly consider the point of view of
Euclidean M-theory on $\calt^3$. The modular group of the torus is
$\Sl3$. There are seven scalar fields that arise from the six moduli of
$\calt^3$
and the value of $C_{ijk}$, the component of the three-form
potential in the
toroidal directions.
Instantons arise from two sources. On the one hand there are three
integers
associated with the Kaluza--Klein modes. On the other hand the
world-volume
of the
membrane can wrap on the torus with winding numbers associated with the
three directions. If the compactification is viewed
in two stages it is related to the $\calt^2$ compactification of type IIA
theory. In the
first stage
consider a single compact dimension which gives the Kaluza--Klein modes
that are $D0$-branes of the type IIA theory. In
addition the wrapped membrane gives fundamental IIA strings -- with
tensions
that are multiples of the fundamental tension according to the
wrapping number.
In the second stage the Euclidean IIA theory is compactified on
$\calt^2$.
The world-line of a charge-$n$ $D0$-brane can wind arbitrarily around
either cycle, giving two
further integers. The world-volume of the fundamental string can
also wrap on
the torus, giving two further integers in addition to the windings of the
membrane around the eleventh dimension.\foot{A very recent preprint
\refs{\tthree} considers Matrix theory on $\calt^3$ which involves some
related issues.} In the limit in which one direction decompactifies only
those configurations with zero winding number in that direction
survive and
the nine-dimensional result should be recovered.

Now consider the point of view of Euclidean IIB on $\calt^2$. We shall
denote the complex structure of the two-torus by
\eqn\compst{U\equiv U_1+iU_2  = {1\over
g^B_{11}} (g^B_{12} + i \sqrt {-g^B}),}
and the K{\"a}hler structure by
\eqn\kahler{T\equiv T_1 + i T_2 = B + i \sqrt {-g^B},}
where $B$ is the component of the \NSNS\ two-form with indices in the
directions of the torus and $g^B_{ij}$ ($i=10,11$) are components of the
string metric
in these two directions. The determinant of this sub-metric is $g^B =
g^B_{10\ 10}
g^B_{11\ 11} - (g^B_{10\ 11})^2 $.
The seven scalar fields are the six real and imaginary parts of
$\rho$, $T$
and $U$, together with the
components of the \RR\
two-form in the directions of the torus, $C^{(2)}$. In the IIB
language the
factor of $\Sl2$ in the duality group $\Sl3 \times \Sl2$ acts on $U$
and none of the other fields.
We shall refer to this as the group $Sl_U(2,{\bf Z})$. The $\Sl2$ groups
associated with the fields $\rho$ and $T$ ($Sl_\rho(2,{\bf Z})$,
$Sl_T(2,{\bf Z})$) are non-commuting subgroups in $\Sl3$. The $Z_2$
transformation $\rho \to - 1/\rho$ in $Sl_\rho(2,{\bf Z})$ induces the
transformations $B\to C^{(2)}$ and $\phi \to -\phi$ (in the
Einstein frame)
so that its action on $T$ is
\eqn\resty{T\to \tilde T, \qquad \tilde T \to T, \qquad U\to U}
where,
\eqn\trans{\tilde T \equiv \tilde T_1 + i \tilde T_2 = C^{(2)} + i
e^{-\phi^B}
\sqrt {-g^B}} is the K{\"a}hler
structure of the D-string torus. The $Z_2$ transformation in $Sl_T(2,{\bf
Z})$, $T\to -1/T$, induces
\eqn\transtwo{\tilde T \to \rho_2, \qquad \rho_2 \to \tilde
T,\qquad U\to U.}

The complete duality-invariant eight-dimensional expression will be
considered in \refs{\greenhove}. Here, we will indicate how those terms
that reduce to \rfours\ in the decompactification limit to the
ten-dimensional IIB theory may be calculated. The dualities of the IIB
theory may be used to map the expression for the one-loop
four-graviton amplitude in fundamental string theory into the
non-perturbative
expression for the $R^4$ term, as follows. The one-loop amplitude
with four
external gravitons can be calculated in eight dimensions in a manner very
similar to that in \refs{\greenbrink, \dkl} using \Loop\ and \latII.
It has the form (in the string frame)
\eqn\exact{{1\over\kappa_{10}^2}{\pi\over3\cdot 2^8}\int d^8 x
\sqrt{-g^{(8)}} \Rfour\left( L(T,\bar T) + L( U, \bar U)\right),}
where
\eqn\ldef{ L(T,\bar T) = \ln (T_2| \eta (T)|^4) ,} and $\eta(T)$ is the
Dedekind function.
This can be expanded into an
infinite series of terms,
\eqn\expand{ L(T,\bar T) \equiv \sum_{m,n=0}^\infty  L_{mn} = \ln(T_2) +
{\pi\over 3} T_2 + \sum_{m,n>0} {1\over n} \left(e^{-2i\pi mn
\bar T}+e^{2i\pi mn T}\right),}
with a similar expansion for the function of $U$.

In writing \exact\ we have subtracted the logarithmic divergence that
arises in the low
energy eight-dimensional theory ($N=8$ supergravity in eight
dimensions) by
imposing the
requirement that it be invariant under $Sl_T(2,{\bf Z})\otimes
Sl_U(2,{\bf
Z})$ transformations. The divergent piece is proportional to the $\beta$
function for the $R^4$ interaction. This is the same procedure as the one
in \dkl.
The presence of the factor $\ln T_2$ reflects this logarithmic
divergence.
The expression \exact\ also contains a term proportional to $T_2$,
which is
necessary for
it to decompactify to the correct ten-dimensional expression. This term
arises from string world-sheets that do not wrap around either
cycle of the
torus. In fact, knowledge of the logarithmic term and the linear term,
together with the $\Sl2$ symmetry and the T-duality symmetry which
takes either of the type II theories into itself in eight dimensions, is
sufficient to determine \exact\
completely. The double sum over $m$ and $n$ comes from configurations in
the functional integral in
which the fundamental string
world-sheet winds around each of the two cycles of the torus a
non-zero number
of times. Upon decompactification ($T_2 \to \infty$) only the
linear $T_2$
term contributes to
the ten-dimensional action. Similarly, the $U$ term in \exact\ can be
interpreted in terms of
degenerate wrappings of world-sheets in which the world-sheet coordinates
wind around a
single cycle of the torus.

This interpretation of the terms in this series can be verified
explicitly
by evaluation of the path
integral for a string wrapped around $\calt^2$ starting with Nambu action
and using the \lq Schild gauge' as in \luc. The Schild gauge is one in
which the action is the square of the Nambu action which is
invariant under
symplectic diffeomorphisms. This will be
described in detail in \refs{\greenhove}.

A $Z_2$ S-duality transformation, $\rho \to - 1/ \rho$, \resty\ converts
the fundamental string (the F-string) to a D-string. This converts the
terms in \expand\ with non-zero windings of the fundamental string, $m,n
\ne
0$, into corresponding terms for the wrapped D-string. These can again be
obtained directly by
functional integration over the wrapped D-string world-sheet,
starting now
with the Euclidean
Dirac--Born--Infeld action,
\eqn\dbiact{\C|L_{DBI}={1\over 2\pi} \int d^2 \xi \left(n e^{-\phi^B}
\sqrt {-\det(G + \C|F)} + {i\over 2} n C^{(2)} + 2in C^{(0)} \wedge
\C|F\right)}
where $\C|F = F - B$ and $F=dA$ is the field strength of the world-volume
vector potential, $A$.
This contains the world-sheets for general D-strings with charges
$(p,n)$.
For the present argument it is sufficient to keep only the terms with
$(0,n)$, which are T-dual to D-instantons. The functional integral now
includes integration over ${\cal F}$ which gives rise to a nontrivial
factor in the measure. The result of the functional integral is that the
non-perturbative terms in \expand\ with $m,n\ne 0$ are replaced by
\refs{\greenhove}
\eqn\bornnew{
\tilde L_{mn} =   e^{2\pi i mn C^{(2)}} \left|m\over n\right|
{e^{-\phi^B}\over \sqrt{e^{-2\phi^B}+(C^{(0)})^2}} K_1\left(2\pi |m|n T_2
\sqrt{e^{-2\phi^B}+(C^{(0)})^2} \right).}

Making the T-duality transformation, \transtwo, replaces $e^{-\phi^B}$ by
$\sqrt{-g^B} e^{-\phi^B}$ in \bornnew and the wrapped D-strings by
D-instantons.
Taking the limit $\sqrt{-g^B} \to \infty$ decompactifies the dual
torus to give
\eqn\bornagain{
\hat L_{mn}  (\rho^B,\bar \rho^B) = (\rho_2^B)^{1/2}e^{2\pi i mn\rho^B_1}
\left|m\over n\right|
K_1\left(2\pi |m|n \rho_2 \right),
}
which agrees with the conjectured ten-dimensional result in
\bessum\ up to
an overall constant whose value depends on a detailed calculation of the
measure.

\newsec{Comments concerning supersymmetry.}

We have presented some evidence that the scalar function  of the moduli
fields multiplying the $R^4$ terms of  M-theory on a torus is
determined by
perturbative and non-perturbative duality symmetries.    The
expression in
\mainres\  reproduces the precise coefficients of the tree-level and
one-loop perturbation theory results in nine-dimensional IIA and IIB
superstring theories,  together with the infinite series of
instanton terms
that are associated with  the expected $\Sl2$ symmetry.  Although we do
not have a proof that these conditions uniquely determine the
function for
arbitrary moduli, the decompactification limits to both of the type II
ten-dimensional string theories, as well as to eleven-dimensional
M-theory,
are uniquely determined if these limits are assumed to be uniform.   The
ten-dimensional IIB limit coincides with the D-instanton sum  conjectured
in \refs{\gg}.  The scalar function  in \mainres\ is   an
eigenfunction of
the Laplace equation  on the fundamental domain of $\Sl2$, \laplaceq,  so
that it is uniquely determined by the tree and one-loop terms in its
expansion.  Proving that this conjectured function is indeed correct
therefore amounts to understanding why it should be an
eigenfunction of the
Laplace equation ---  a condition that should follow from the
constraints of
supersymmetry.  The fact that the coefficient of the $\Rfour$ term should
be determined by
supersymmetry even though it  is a non-holomorphic function of the moduli
and contributes to two terms in the perturbation expansion  is  rather
unusual.

One consequence of this structure is that  there should be a
non-renormalization theorem in either of the type II string theories that
prevents perturbative contributions to the $R^4$ terms beyond one
loop and
prevents non-perturbative contributions to the IIA theory in ten
dimensions.
Furthermore,  the coefficient of the M-theory $R^4$ term in eleven
dimensions is determined by the coefficient of the one-loop term in
either
of the ten-dimensional type II theories, as in \mresult.
Gratifyingly,  this
eleven-dimensional expression has an independent motivation based on
supersymmetry in ten dimensions.
This can be seen to follow from its relation to the term in the M-theory
effective action that is an eleven-form $C^{(3)}\wedge X_8$
\refs{\duffminas} which is known to be present from a variety of
arguments,
such as anomaly cancellation \refs{\horavawit}. The expression
$X_8$ is the
eight-form in the curvatures that is inherited from the term in type IIA
superstring theory \refs{\vafawitten} which is given by
\eqn\vafwit{-   \int {d^{10}x \over  (2\pi)^5 }  B\wedge  X_8 =  -  
{1\over 2}
\int  {d^{10}x \over  (2\pi)^5 } \sqrt{-g^{A(10)}}\epsilon_{10} B X_8.}
where,\foot{Here,
$R^n$ is the outer product of $n$ Riemann curvatures where $R$ is
viewed as
a two-form matrix in the $10\times 10$-dimensional representation of
$SO(9,1)$. } \eqn\eightform{X_8 = {1 \over 192} \left(\tr R^4 -
{1\over 4}
(\tr R^2)^2\right).}
This is consistent with the antisymmetric tensor gauge symmetry, as
can be
seen by the replacement $B \to B + d\Lambda^{(1)}$ and an integration by
parts.

To see how this term is related to the $\Rfour$ term discussed in this
paper recall first that  in the case of the heterotic and type I
superstrings, ten-dimensional $N=1$ supersymmetry provides powerful
constraints on terms that are related to parity-violating
anomaly-cancelling terms \refs{\lerche}.   An example  of the power of
these constraints is the explicit determination of  highly nontrivial
non-perturbative  relationships between heterotic $SO(32)$ and type I
theories in eight dimensions  \refs{\bfkov}.
These strong constraints follow from the structure of the two independent
ten-dimensional $N=1$
super-invariants which contain an odd-parity term
\refs{\lerche,\roo,\tseytlin}
,
\eqn\supthree{I_3= t_8 \tr R^4 - {1\over 4} \epsilon_{10}B \tr R^4 } and
\eqn\supfour{I_4= t_8 (\tr R^2)^2 - {1\over 4} \epsilon_{10}B (\tr
R^2)^2}
(where the notation is that of \refs{\tseytlin}). Using the fact that
$t_8t_8R^4=24t_8tr(R^4)-6t_8(trR^2)^2$, it follows that  the
particular linear
combination,
\eqn\lincomb{I_3 - {1\over 4}I_4 = {1\over 24} t_8 t_8 R^4 - 48
\epsilon_{10}B\ X_8}
contains both the ten-form $B\wedge X_8$ and $t_8 t_8 R^4$.
Therefore, the
part of the effective action that contains \vafwit\ must be
\eqn\sprimedef{S'_{R^4} = {1\over\kappa_{10}^2}{\pi^2\over 48 } \int
d^{10}x \sqrt{-g^{A(10)}} ( I_3 -{1\over 4} I_4) } so that $S_{R^4}'$
contains precisely the torus contribution to the $  t_8   t_8 R^4$
term in
\mact. In other words, the torus contribution to the $  t_8  t_8
R^4$ term
in the IIA theory combines with the ten-form in the linear combination
\lincomb\ which are bosonic terms in an $N=2$ super-invariant. At strong
coupling this lifts to a super-invariant of the eleven-dimensional theory
in which the $\Rfour$ terms have the same coefficient as in \exact.

At strong coupling this lifts to the $\Rfour$ terms in eleven  
dimensions which  have the same coefficient as in \exact.  The full  
supersymmetric  effective action must therefore  contain these terms  
(along with others that we have not considered here).

The other $\Rfour $ terms in
\mact\ depend non-trivially on the dilaton, $\rho_2^A$, whereas no
dilaton
dependence can be introduced into the $B\wedge X_8$ term without spoiling
the antisymmetric tensor gauge symmetry. For this reason neither the
tree-level $t_8 t_8 R^4$ term in the effective ten-dimensional IIA
action,
nor the instanton corrections in lower dimensions, can be related to the
ten-form in an obvious manner.
However, once terms beyond the lowest-order terms in the effective action
are known supersymmetry probably determines the whole action.

Our interest in this subject is linked to the related question of whether
the $F^4$ terms that enter into the description of $D0$-brane scattering
\refs{\swedes,\kabat,\douglasetal} are renormalized. This question has a
direct connection with the issues discussed in this paper due to another
set of duality relations.\foot{This argument was formulated in
collaboration with Constantin Bachas and will be presented in detail
elsewhere.}     Firstly, upon
compactification on a circle to nine dimensions the process in which two
$D0$-branes scatter is mapped by T-duality into the scattering of two
D-strings with unit winding number around the compact direction in the
Einstein frame. This process is related in turn by S-duality for the type
IIB theory into the scattering of two fundamental strings (\lq
F-strings')
that are also wound around the circle and are BPS states carrying no
momentum in the tenth dimension. This scattering amplitude can be
calculated directly to any order in string perturbation theory. In
order to
make this sequence of statements precise there are delicate questions
concerning the mapping between the configuration appropriate for the
scattering of $D0$-branes at a fixed impact parameter and the
scattering of
F-string states at fixed momentum transfer.

\vskip 0.5cm
{\it Acknowledgements:} We are grateful to Costas Bachas and Elias
Kiritsis for
useful conversations and encouragement. P. V. is grateful to the Ecole
Polytechnique and DAMTP for financial support and EC support under the
Human Capital and Mobility programme is also gratefully acknowledged.

\listrefs
\end